\documentclass[]{article}
\usepackage{graphicx}

\begin{document}

\centerline{\Large \bf Chromosome-Length-Scaling in}\smallskip
\centerline{\Large \bf Haploid, Asexual Reproduction}\medskip

\bigskip

\centerline{P.M.C. de Oliveira}

\bigskip

Instituto de F\'\i sica\par

Universidade Federal Fluminense\par

av. Litor\^anea s/n, Boa Viagem, Niter\'oi, Brasil 24210-340

\medskip

e-mail addresses: pmco@if.uff.br

\bigskip

\begin{abstract}

        We study the genetic behaviour of a population formed by haploid
individuals which reproduce asexually. The genetic information for each
individual is stored along a bit-string (or chromosome) with $\, L\, $ bits,
where $\, 0$-bits represent the wild-type allele and $\, 1$-bits correspond to
harmful mutations. Each newborn inherits this chromosome from its parent with
some few random mutations: on average a fixed number $\, m\, $ of bits are
flipped. Selection is implemented according to the number $\, N\, $ of $\,
1$-bits counted along the individual's chromosome: the smaller $\, N\, $ the
higher the probability an individual has to survive a new time step. Such a
population evolves, with births and deaths, and its genetic distribution
becomes stabilised after many enough generations have passed.

        The question we pose concerns the procedure of increasing $\, L\, $. 
The aim is to get the same distribution of relative genetic loads $\, N/L\, $
among the equilibrated population, in spite of a larger $\, L\, $. Should we
keep the same mutation rate $\, m/L\, $ for different values of $\, L\, $? The
answer is {\sl yes}, which intuitively seems to be plausible. However, this
conclusion is not trivial, according to our simulational results: the question
involves also the population size.

\end{abstract}

\newpage
 \section{Introduction}

        A natural way for evolution is to increase the chromosome length in
order to make space for more genetic information. However, Nature seems to
impose a maximum possible chromosome length, as inferred from at least two
general observations. First, all big animals present the same order of
magnitude for their chromosome lengths. Second, the genetic information for
these animals is spread over more-than-one chromosome pairs (23 for humans),
instead of a single, long one. This broken-information storage strategy demands
an extra cost, a coordinated regulatory mechanism triggering the simultaneous
copy of the various chromosomes at reproduction or cell division. Therefore,
some unavoidable obstacle should exist, which prevents Nature to follow the
simpler rule of increasing the length of a single chromosome.

        On the other hand, during reproduction, the chemical machinery which
performs DNA duplication works as a zipper, scanning the one dimensional
chromosome chain, basis after basis. Thus, the total number $\, m\, $ of
``errors'' (i.e. mutations appearing in offspring as compared to parents)
should be proportional to the chromosome length $\, L\, $, on average.
Therefore, by increasing $\, L\, $, this linear behaviour indicates that one
should keep the same mutation rate $\, m/L\, $: this procedure is then supposed
to yield the same genetic quality for the whole population, in spite of the
larger $\, L\, $.

        In order to test these ideas, we decided to simulate on computers a
very simple model where each individual carries a single chromosome represented
by a bit-string with $\, L\, $ bits. Each bit can appear in either of two
forms, $\, 0\, $ or $\, 1\, $. The allele represented by a $\, 0$-bit is the
wild type, whereas a $\, 1$-bit corresponds to some harmful mutation. In this
work, we treat the case of haploid individuals, the genetic information stored
along a single bit-string. In another related work \cite{sex} we treat diploid,
sexual reproducing populations with crossings and recombination, and the
results and conclusions are completely different (preliminary results can be
found in \cite{newbook}).

        The model is designed in order to keep only the fundamental ingredients
of genetic inheritance and Darwinian evolution: random mutations performed at
birth, and natural selection. All other biological issues which have no direct
influence during the very moment of reproduction are ignored, as, for instance,
the various correlations and inhomogeneities along the chromosome, the various
phases during embryo development, the metabolism during life, etc. Instead,
selection is implemented by taking into account just one phenotype: the number
$\, N\, $ of harmful alleles, i.e. $1$-bits counted along the individual's
chromosome. It is a {\sl minimalist} model, not a {\sl reductionist} one,
because we do not divide the problem into smaller, separate pieces neither in
space nor in time (see \cite{parisi}). This distinction between minimalism and
reductionism is of fundamental importance in what concerns the size- and
time-scaling behaviour which leads to the criticality observed in evolutionary
systems \cite{newbook}.

        We keep in the computer memory the chromosome (bit-string) of each
individual belonging to a population (in reality, the number of 1-bits is
enough). The total number $\, P\, $ of individuals is kept fixed by controlling
the number of deaths equal to the number of births at each new time step: a
fraction $\, b\, $ of the population dies, while another fraction also equal to
$\, b\, $ of newborns are generated by random parents chosen among the
survivors. At each time step, we perform two successive sub-steps, first deaths
and then births.

        The death sub-step contains the selection ingredient \cite{tb}: an
individual with $\, N+1\, $ harmful alleles ($\, 1$-bits) along its chromosome
survives with a smaller probability than another individual with $\, N\, $
harmful alleles. Let's call $\, x\, $ the ratio between these two
probabilities, $\, x = P(N+1)/P(N)\, $, and let's also consider $\, x\, $
independent of $\, N = 0,\, 1,\, 2 \dots L\, $ (this assumption is equivalent
to adopt an exponential decay for the survival probability as a function of $\,
N\, $). The same $\, x\, $ is also adopted as the survival probability for
individuals with $\, N = 0\, $. Of course, the value of $\, x\, $ should be
strictly smaller than 1, otherwise nobody dies and the population does not
evolve.

        Let's now describe the first sub-step corresponding to deaths. First,
we count the number $\, H(N)\, $ of individuals with $\, N\, $ harmful alleles. 
Then, we solve the polynomial equation

$$\sum_{N=0}^L\, H(N)\, x^{N+1}\, =\, (1-b)\, P\eqno(1)$$

\noindent getting the value of $\, x\, $. We adopted $\, b = 0.02\, $, i.e. 
$2\%$ of the population die each new time step. Other not-too-large values
($1\%,\, 3\%\, $, etc) can also be adopted with the same results, since the
role of $\, b\, $ is only to fix a convenient time scale, the time interval
between two successive snapshots of a movie describing the evolving population.
As $\, b\, $ is a small fraction, $\, x\, $ is always near $\, 1\, $ (in fact,
$\, 1-b \le x < 1\, $). Now, for each individual $\, i\, $, we toss a random
number $\, r\, $ in between $\, 0\, $ and $\, 1\, $: if $\, r < x^{N_i+1}\, $,
then this individual survives, where $\, N_i\, $ counts the number of harmful
alleles along its chromosome; otherwise, this individual dies and is excluded
from the population. After applying this death roulette to the whole
population, the number of survivors is $\, (1-b)\, P\, $, on average.

        The second sub-step corresponds to births in exactly the same number of
deaths occurred during the first. For each newborn we toss a random parent
among the survivors. Then, we copy its chromosome and perform mutations on the
copy. Each mutation is a bit which is flipped (from $\, 0\, $ to $\, 1\, $, or
vice-versa), the position of which along the chromosome taken at random.  As
the number of $\, 0$-bits is dominant among the population, ``bad'' mutations
($\, 0\, $ to $\, 1\, $) are much more likely to occur, as in Nature. The total
number of mutations for this particular newborn is a random number $\, M\, $
whose average coincides with the parameter $\, m\, $ fixed the same for the
whole population during all the evolutionary time: for each newborn, we toss
$\, M\, $ in between $\, 0\, $ and $\, 2m\, $. Neither $\, M\, $ nor $\, m\, $
need to be integer numbers, they are real numbers. Suppose the tossed $\, M\, $
is not an integer. Then, we perform first $\, {\rm int}(M)\, $ mutations, where
$\, {\rm int}(X)\, $ is the integer part of $\, X\, $. After that, with
probability $\, {\rm frac}(M)\, $, we perform a last mutation, where $\, {\rm
frac}(X)\, $ is the fractional part of $\, X\, $: we toss a new random number
$\, r\, $ in between $\, 0\, $ and $\, 1\, $, and perform the last mutation
only if $\, r < {\rm frac}(M)\, $.

\begin{figure}[!hbt]

\begin{center}
 \includegraphics[angle=-90,scale=0.5]{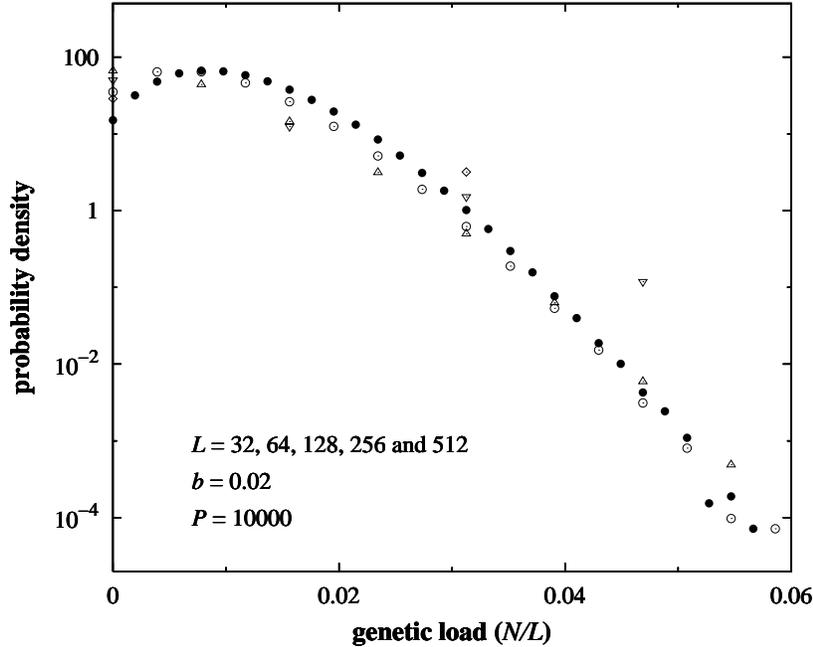}
\end{center}

\caption{Collapsed distributions of the individual genetic load $\, N/L\, $,
for haploid, asexual reproducing populations with different chromosome lengths,
after many generations. The full circles correspond to the largest length $\, L
= 512\, $. The mutation rate $\, m/L = 1/320 \approx 0.003\, $ is the same for
all lengths, as well as the population size $\, P = 10000\, $. The inverse of
the typical genetic load (here, $\, \langle N\rangle/L \approx 0.01\, $) is an
estimate for the population genetic quality: the larger $\, \langle N\rangle/L
\, $ the poorer this quality.}

\label{fig1}
\end{figure}

        The simulation starts with all individuals alike, only $\, 0$-bits
along their chromosomes, i.e. $\, N = 0\, $ for all. As the generations pass,
individuals with different values of $\, N = 1, 2, 3 \dots\, $ appear, due to
mutations. After many generations, the distribution of genetic loads $\, N/L\,
$ stabilises. Fig.1 is an example, where we have superimposed different
chromosome lengths. The first observation concerning this figure is that both
Darwinian evolution ingredients (random mutations and selection) work together.
In order to better understand this important point, imagine the first sub-step
was replaced by random deaths (no selection): in this case, the curve would run
away to the right, sticking to a normal, bell-shaped narrow distribution
(Gaussian) centred in $\, N/L = 0.5\, $, far to the right in Fig.1. Moreover,
in this selection-less case the wild-type genotype $\, N = 0\, $ would be
extinct. This would correspond to a completely random genetic pool, nothing to
do with any kind of evolutionary process. On the other hand, instead of
selection, imagine we skip the mutation ingredient (no mutations): now, the
curve would be replaced by a single point at $\, N = 0\, $, a situation where
all individuals are ``perfect'', again no relation with any evolutionary
process. Without mutations, this would be the final destiny even if we have
started the simulation from a randomly chosen population. The fact that Fig.1
is in between these two extremes, neither $\, N = 0\, $ nor $\, \langle
N\rangle/L = 0.5\, $, shows that Darwin evolution is going on, the tendency
towards complete genetic randomisation due to successive mutations is
compensated by the selective deaths, according to a steady-state balance. In
the physicist's jargon, we can say that selection is able to contain the
tendency towards entropy explosion, the same balance which leads to free-energy
minimisation.

        The second observation concerning this Fig.1 is that all populations
with different chromosome lengths collapse onto the same curve. In other words,
one is able to obtain the same genetic quality for populations with different
chromosome lengths, provided one keeps the same mutation rate $\, m/L\, $ per
locus.

        Based on these results, the preliminary conclusion would be the
following. There is no price to pay by adopting the evolutive procedure of
increasing the chromosome length. One can perform this increment by keeping the
same chemical copying machinery, i.e. the same error rate $\, m/L\, $, and
obtain the same degree of genetic degradation kept under control. The advantage
is a larger information storage capacity. Why, then, real chromosome lengths
seem to have already reached a limiting size?

        The story is incomplete. The rest is described in the following
section. Definitive conclusions appear in the last section.

\newpage
 \section{The Scaling}

        Fig.1 with chromosome lengths of $\, L = 32,\, 64,\, 128\, $ and 512 is
incomplete. By including a larger length of $\, L = 1024\, $, one gets Fig.2.

\begin{figure}[!hbt]

\begin{center}
 \includegraphics[angle=-90,scale=0.5]{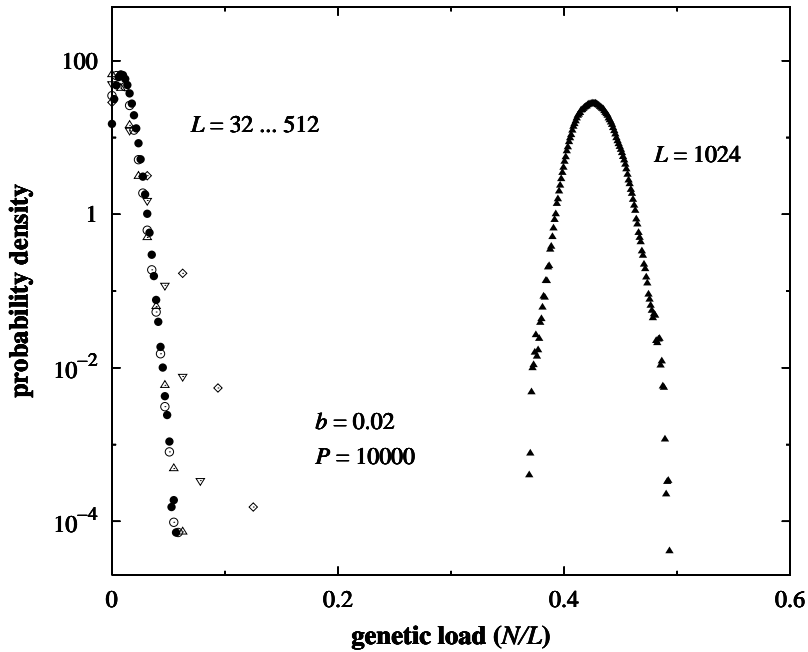}
\end{center}

\caption{The same data displayed in Fig.1, now within a wider horizontal scale
in order to fit further data obtained for a larger chromosome length of $\, L =
1024\, $ (rightmost bell shaped curve). The population size is still $\, P =
10000\, $, the birth/death rate $\, b = 0.02\, $, and $\, m/L = 1/320 \approx
0.003\, $.}

\label{fig2}
\end{figure}

        Unlike the collapsed curves of Fig.1, now repeated at the left-handed
side of Fig.2, the larger chromosome length of $\, L = 1024\, $ shows a runaway
from the wild-type genotype ($\, N = 0\, $) towards the random situation ($\,
\langle N\rangle \approx L/2\, $). Beyond this length, the wild-type genotype
is extinct and the whole population distribution is no longer glued to it. The
same behaviour is also observed in many other similar systems, in particular
the pioneering Eigen model \cite{eigen}. Beyond a certain limit for the
chromosome length $\, L\, $, the scaling properties denoted by the collapse of
many curves into a single distribution containing the wild-type genotype no
longer hold. The preliminary conclusion at the end of last section is now in
check. We need a more detailed analysis, which follows.

        Let's consider different values for the parameter $\, m\, $, the
average number of mutations performed at birth. Fig.3 shows the average genetic
load $\, \langle N\rangle/L\, $ as a function of $\, m\, $, for various
chromosome lengths $\, L = 32,\, 64,\, 128\, \dots\, $ 2048 (black circles) and
4096 (black squares). In the limit of large enough chromosome lengths, this
figure seems to display a first order phase transition. The average genetic
load vanishes if the number $\, m\, $ of mutations remains below a certain
threshold $\, m_{\rm c}\, $ (here, $\, m_{\rm c} \approx 3\, $). This would be
the survival phase, where the population genetic quality is not compromised by
too much mutations at birth. On the other hand, for $\, m > m_{\rm c}\, $, one
observes the average genetic load suddenly jumping up and approaching the pure
random value $\, \langle N\rangle/L = 0.5\, $, where no evolution is possible,
as explained in the next two paragraphs.

\begin{figure}[!hbt]

\begin{center}
 \includegraphics[angle=-90,scale=0.5]{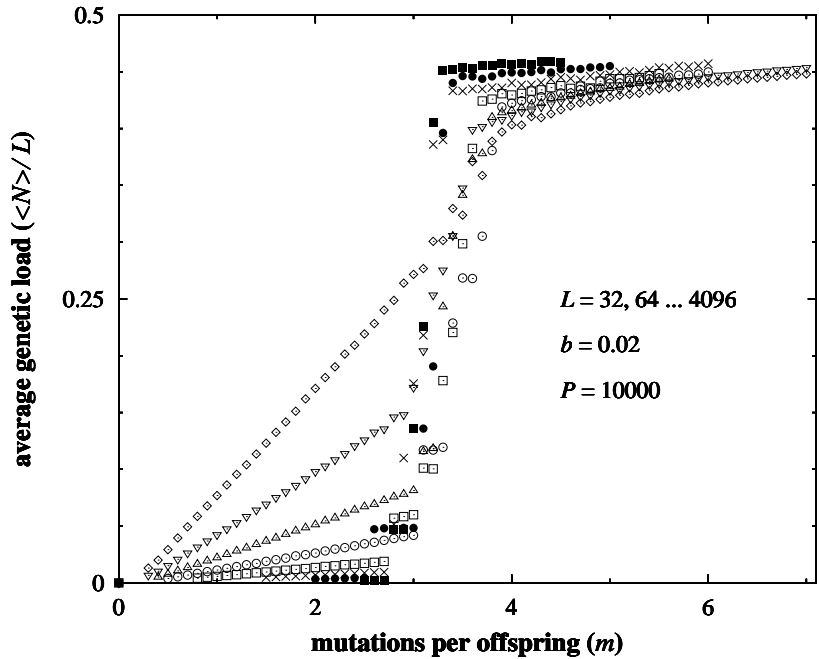}
\end{center}

\caption{Apparent first order phase transition: survival on the left,
extinction on the right. For large enough chromosome lengths, the curve
approaches a step. Evolution is possible only below this step, on the
left-handed side, where the genetic degradation (as measured by the density of
``bad genes'' read on the vertical axis) is far below the random behaviour $\,
\langle N\rangle/L \approx 0.5\, $. The number $\, m\, $ of mutations performed
at birth should be smaller than the threshold $\, m_{\rm c} \approx 3\, $.
Beyond this point, the sudden jump towards the random behaviour forbids
evolution to proceed.}

\label{fig3}
\end{figure}

        After the runaway observed in Fig.2 for $\, L = 1024\, $, the
distribution curves for larger and larger values of $\, L\, $ (not shown in
Fig.2 for clarity) would be sharper and sharper, all of them centred near $\,
N/L \approx 0.5\, $, reaching $\, N/L\, = 0.5\, $ for $\, L \to \infty\, $. 
Therefore, within negligible (sub-linear) fluctuations, all individuals share
the same phenotype $\, N \approx L/2\, $ and become selectively alike to each
other. No selection, no evolution.

        Technically, by putting such a sharp distribution $\, H(N)\, $ in
Eq.(1) one gets the solution $\, x = 1\, $ in the limit of large values of $\,
L\, $ (or $\, N\, $). However, this solution is a biological nonsense, because
it implies eternal survival for all individuals, again no evolution. In
reality, for such sharp distributions far from $\, N/L = 0\, $, the population
would undergo a genetic meltdown, and will be eventually extinct. This fate is
artificially avoided by our assumption of a constant size population, which no
longer holds.  We could correct this failure and observe real extinction,
simply by imposing some maximum value $\, x_{\rm max}\, $ near but strictly
smaller than unity, if the solution $\, x\, $ obtained from Eq.(1) surpasses
this limit. However, this procedure is unnecessary because we are interested
only in the survival which holds on the left-handed side of Fig.3, where always
one gets $\, x < 1\, $ from Eq.(1).

\begin{figure}[!hbt]

\begin{center}
 \includegraphics[angle=-90,scale=0.5]{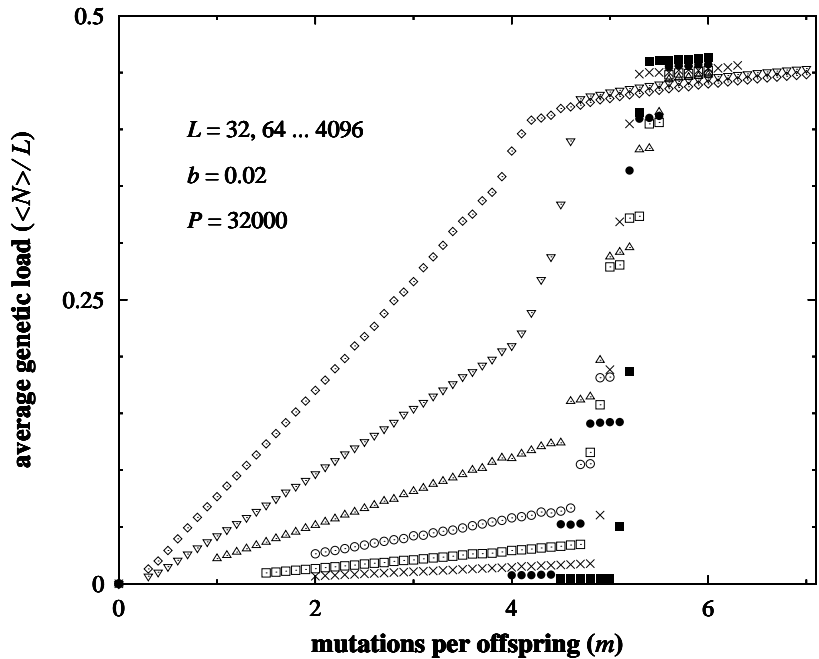}
\end{center}

\caption{For a larger population of $\, P = 32000\, $, the apparent transition
occurs at a larger threshold $\, m_{\rm c} \approx 5\, $, as compared to
Fig.3.}

\label{fig4}
\end{figure}

        On the survival side of Fig.3 or Fig.4, the plots correspond to
straight lines starting at the origin, whose slopes decrease with increasing
$\, L\, $ proportionally to $\, 1/L\, $. Thus, by keeping the same ratio $\,
m/L\, $ for increasing values of $\, L\, $, one gets always the same value for
$\, \langle N\rangle/L\, $ read on the vertical axis along a plateau, provided
$\, m\, $ does not surpass the transition point $\, m_{\rm c}\, $ (i.e.
provided $\, L\, $ is not too large). In reality, not only the average $\,
\langle N\rangle/L\, $, but the whole distribution of $\, N/L\, $ does not
depend on $\, L\, $, as in Fig.1.

        Is this phase transition genuine? In order to answer this question, we
need to consider the so-called thermodynamic limit (the limit of larger and
larger populations) and ask whether the (would-be) phase transition remains. 
According to traditional statistical physics, phase transitions occur only in
this limit. We could, for instance, repeat Fig.1 for a larger population, say
$\, P = 32000\, $. We do not need to show such a plot, because it is the same
as Fig.1. The only difference is that $\, L = 1024\, $ now fits into the same
collapsed curve shown in Fig.1, instead of following the runaway observed in
Fig.2. In fact, the runaway does not occur if the population size is large
enough. Fig.4 corresponds to a larger population of $\, P = 32000\, $, to be
compared with the former Fig.3: now, the apparent transition point is located
at $\, m_{\rm c} \approx 5\, $, larger than the former value.

\begin{figure}[!hbt]

\begin{center}
 \includegraphics[angle=-90,scale=0.5]{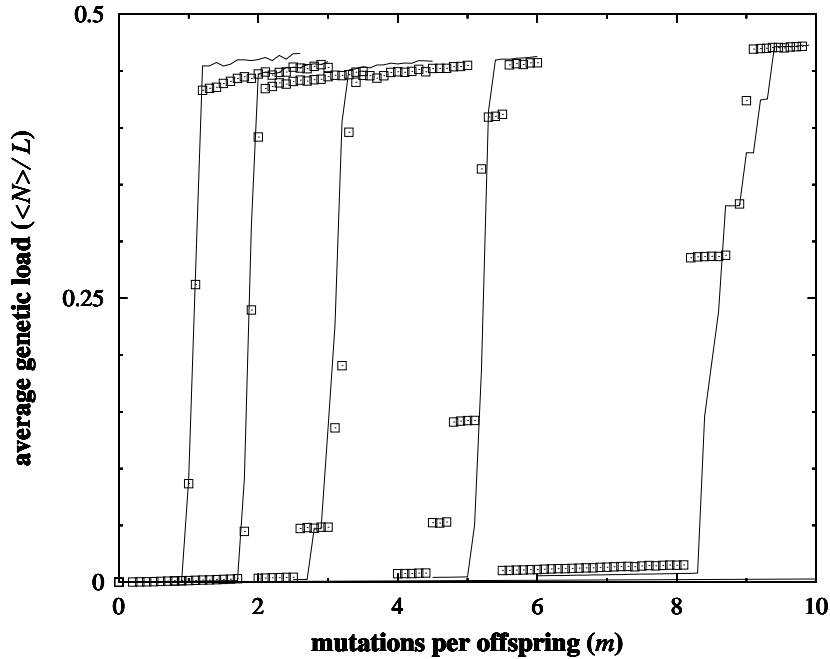}
\end{center}

\caption{The (would-be) transition occurs at different locations for different
population sizes $\, P = 1000,\, 3200,\, 10000,\, 32000\, $ and 100000 from
left to right. Squares correspond to $\, L = 2048\, $, and lines to $\, L =
4096\, $.}

\label{fig5}
\end{figure}

        Fig.5 shows again the average genetic load $\, \langle N\rangle/L\, $
as a function of $\, m\, $, for increasing populations sizes. For clarity, only
data corresponding to the two largest chromosome lengths $\, L = 2048\, $
(squares) and 4096 (lines) are shown. The larger the population size, the
larger the transition point $\, m_{\rm c}\, $. As an estimate for $\, m_{\rm
c}\, $, we have taken the crossings of the $\, L = 4096\, $ curves with the
horizontal line $\, \langle N\rangle/L = 0.25\, $ (just half-way from the
complete order $\, N = 0\, $ towards the complete disorder $\, \langle N\rangle
= L/2\, $). The resulting values of $\, m_{\rm c}\, $ obtained from these
crossings are plotted against $\, P\, $ in Fig.6. They follow a power-law, and
this behaviour indicates that $\, m_{\rm c}\, $ grows indefinitely for larger
and larger populations. In this limit $\, P \to \infty\, $ at fixed $\, m/L\,
$, only the survival phase exists, no runaway.

\begin{figure}[!hbt]

\begin{center}
 \includegraphics[angle=-90,scale=0.5]{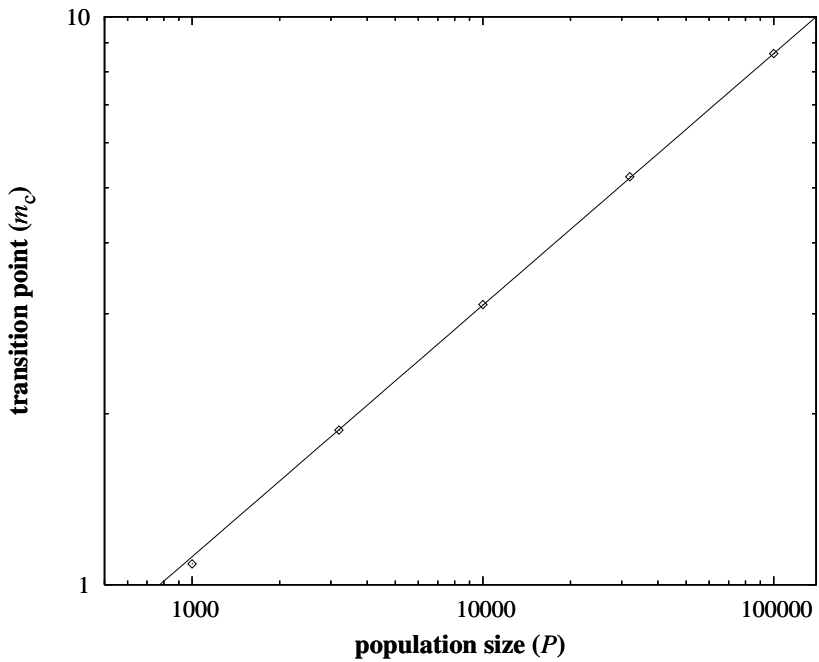}
\end{center}

\caption{The transition point increases for increasing population sizes,
following a power-law. The straight line $\, m_{\rm c} \propto P^{0.44}\, $
fits very well the data. It means that one needs a minimum population size $\,
P_{\rm min} \propto L^{1/0.44} = L^{2.3}\, $ in order to sustain the population
survival at a fixed $\, m/L\, $, where $\, L\, $ is the chromosome length.}

\label{fig6}
\end{figure}

        The apparent survival-extinction transition shown in plots like Fig.3
is not a genuine phase transition, it disappears for large enough population
sizes. Therefore, the preliminary conclusion we have stated at the end of last
section is not completely wrong. Indeed, in order to keep the same genetic
quality of the population for increasing chromosome lengths, one should keep
the same rate of errors when each chromosome is copied for reproduction, i.e. 
the same probability of error {\sl per locus}, $\, m/L\, $. However, this is
not a priceless procedure. The population size should also be large enough to
avoid the runaway shown in Fig.2. The minimum required population size depends
on the largest chromosome length one wants to reach.

        The current paper deals with asexual populations. In another paper
\cite{sex} we have shown that this conclusion does not hold for sexually
reproducing populations. In this case, the survival-extinction transition is a
genuine one, the transition point $\, m_{\rm c}\, $ does not depend on the
population size. When the chromosome length is increased, one should keep the
same {\sl absolute} number $\, m\, $ of mutations performed at birth, not the
ratio $\, m/L\, $. Therefore, with sex, the price to pay for increasing the
chromosome length is higher: one should improve the performance of the chemical
DNA copying machinery, in order to keep the same number of errors, in spite of
a larger length to be copied.

        We close this section with a technical comment on the simulations. The
evolutionary time one needs to reach genetic stabilisation is very huge,
particularly near the jumps shown in plots like Fig.3, Fig.4 or Fig.5. Starting
from an initial population with only $\, N = 0\, $ individuals, one needs to
evolve the whole population through $\, 10^8\, $ or more time steps in order to
reach the runaway shown in Fig.2. In order to control the statistics, we
simulated a total of 10 independent populations, taking averages at the end,
with error bars. For that reason, our computer program runs for a long time in
an Athlon/Opteron 250 processor, up to $\approx 10$ days for each point shown
on the rightmost curve on Fig.5.

\newpage
 \section{Conclusions}

        Based on a very simple model which nevertheless contains both
fundamental ingredients which drive Darwin's evolution, namely random mutations
performed at birth and natural selection, we have discovered some chromosome
length scaling properties. The overall result (valid in the thermodynamic
limit) is very simple to state: in order to increase the chromosome length $\,
L\, $, one should keep the same mutation rate $\, m/L\, $ {\sl per locus},
where $\, m\, $ is the average number of point mutations performed at birth.
This result is also in complete agreement with human intuition, since the
number $\, m\, $ of errors found in a chromosome copy should be proportional to
its length $\, L\, $, when performed by the same chemical DNA copying
machinery.

        However, the issue is not so simple, not so intuitive. The validity of
this linear behaviour depends on the population size. It should be large enough
to avoid the genetic meltdown characterised by the runaway shown in Fig.2,
which means extinction. We have also shown that a minimum population size is
required to avoid this, which increases for larger and larger chromosome
lengths as

$$P_{\rm min} \propto L^\alpha$$

\noindent where our numerical estimate for the exponent is $\, \alpha \approx
2.3\, $.

        By including sex with crossings and recombination, another completely
different scaling holds, independent of the population size: $\, m\, $ instead
of $\, m/L\, $ should be preserved when a $\, L-$scaling transformation is
performed \cite{sex,newbook}.

\end{document}